\documentclass[a4paper,fleqn,usenatbib,useAMS]{mnras}

\usepackage{graphicx}	% Including figure files
\usepackage{amsmath}	% Advanced maths commands
\usepackage{multicol}        % Multi-column entries in tables
\usepackage{bm}		% Bold maths symbols, including upright Greek
\usepackage{pdflscape}	% Landscape pages
\usepackage{verbatim}

\usepackage{lineno}
\usepackage[T1]{fontenc}
\usepackage{ae,aecompl}

\usepackage{newtxtext,newtxmath}
\title[GWs from SN Mass Loss and Natal Kicks]{Gravitational Waves from Supernova Mass Loss and Natal Kicks in Close Binaries}

\author[A. M. Holgado and P. M. Ricker]{A. Miguel Holgado$^{1}$\thanks{Contact e-mail: \href{mailto:holgado2@illinois.edu}{holgado2@illinois.edu}}
and
Paul M. Ricker$^1$
\\
$^{1}$Department of Astronomy and National Center for Supercomputing Applications, University of Illinois at Urbana-Champaign, Urbana, IL 61801, USA
}

\date{Last updated \today.}

\pubyear{2015}

\begin{document}
\label{firstpage}
\pagerange{\pageref{firstpage}--\pageref{lastpage}}
\maketitle

\begin{abstract}
Some fraction of compact binaries that merge within a Hubble time may have formed from two massive stars in isolation.
For this isolated-binary formation channel, binaries need to survive two supernova (SN) explosions in addition to surviving common-envelope evolution.
For the SN explosions, both the mass loss and natal kicks change the orbital characteristics, producing either a bound or unbound binary. 
We show that gravitational waves (GWs) may be produced not only from the core-collapse SN process, but also from the SN mass loss and SN natal kick during the pre-SN to post-SN binary transition. 
We model the dynamical evolution of a binary at the time of the second SN explosion with an equation of motion that accounts for the finite timescales of the SN mass loss and the SN natal kick. 
From the dynamical evolution of the binary, we calculate the GW burst signals associated with the SN natal kicks.
We find that such GW bursts may be of interest to future mid-band GW detectors like DECIGO.
We also find that the energy radiated away from the GWs emitted due to the SN mass loss and natal kick may be a significant fraction, ${\gtrsim}10\%$, of the post-SN binary's orbital energy. 
For unbound post-SN binaries, the energy radiated away in GWs tends to be higher than that of bound binaries. 
\end{abstract}


%%%%%%%%%%%%%%%%%%%%%%%%%%%%%%%%%%%%%%%%%%%%%%%%%%

%%%%%%%%%%%%%%%%% BODY OF PAPER %%%%%%%%%%%%%%%%%%

% The MNRAS class isn't designed to include a table of contents, but for this document one is useful.
% I therefore have to do some kludging to make it work without masses of blank space.
\begingroup
\let\clearpage\relax
%\tableofcontents
\endgroup
\newpage

\section{Introduction}
The Laser Interferometer Gravitational-Wave Observatory (LIGO) detection of gravitational waves (GWs) from the binary black-hole (BBH) merger GW150914 marked the beginning of GW astronomy \citep[][]{ligo_scientific_collaboration_and_virgo_collaboration_observation_2016}.
The LIGO-Virgo GW detection of the binary neutron star (BNS) merger GW170817 \citep[][]{abbott_gw170817:_2017} along with the multi-messenger electromagnetic follow-up marked the beginning of multi-messenger astronomy with GWs.
For an isolated-binary formation model, the GW measurements of the masses and the EM measurements of the optical counterpart's offset in the host galaxy NGC4993 \citep[e.g.,][]{coulter_swope_2017,hallinan_radio_2017,kasliwal_illuminating_2017} have provided constraints on the progenitor at the time of the second supernova \citep[SN;][]{abbott_progenitor_2017}. 
\par
SN mass loss and natal kicks are important binary stellar-evolution processes in the formation of compact binaries including BBHs, neutron-star-black-hole (NSBH) binaries, and BNSs that merge within a Hubble time \citep[e.g.,][]{hills_effects_1983,janka_neutron_1994,brandt_effects_1995,kalogera_orbital_1996,fryer_double_1997,colpi_formation_2002,scheck_pulsar_2004,martin_supernova_2009,janka_natal_2013,tauris_formation_2017,wysocki_explaining_2018,michaely_supernova_2018,kochanek_stellar_2019}. 
A variety of mechanisms have been proposed to explain the uncertain origin of SN natal kicks.  
These include hydrodynamical momentum transfer \citep[e.g.,][]{burrows_pulsar_1996,nordhaus_theoretical_2010,nordhaus_hydrodynamic_2012}, asymmetric neutrino emission with and without the presence of magnetar-level magnetic fields \citep[e.g.,][]{kusenko_pulsar_1996,kusenko_pulsar_2004,socrates_neutrino_2005,fryer_effects_2006,sagert_pulsar_2008}, and anisotropic gravitational tugs on the remnant from the asymmetric ejecta distribution \citep[e.g.,][]{janka_neutron_2017}. 
\par
In order to explicitly model compact-object natal kicks in close binaries, global 3-D general-relativistic radiation magnetohydrodynamics simulations of core-collapse SNe with detailed neutrino transport and realistic equations of state for all elemental species are required.
These simulations push the limits of computational feasibility. 
Some have attempted to model the accumulation of a natal kick in hydrodynamical SN simulations \citep[e.g.,][]{scheck_multidimensional_2006,wongwathanarat_hydrodynamical_2010,wongwathanarat_three-dimensional_2013,gessner_hydrodynamical_2018,muller_multidimensional_2018,muller_three-dimensional_2019,nakamura_long-term_2019}. In state-of-the-art binary population synthesis (BPS) codes, the SN process is treated as an instantaneous process and is modeled analytically.
Here, we implement an extended model in order to treat the separate timescales of the SN mass loss and the SN natal kick and compute the corresponding GW emission from this process. 
\par
The evolution of binary massive stars and the formation of compact binaries are still highly uncertain, though observations of populations such as X-ray binaries, Galactic binary pulsars, and compact binary mergers are continually providing constraints on binary-evolution models.
The BPS method provides a framework for interpreting the population properties of compact binaries and has been used for these binary populations \citep[e.g.,][]{belczynski_effect_1999,hurley_evolution_2002,oshaughnessy_constraining_2005,oslowski_population_2011,dominik_double_2012,fragos_x-ray_2013,tzanavaris_modeling_2013,beniamini_natal_2016,bray_neutron_2016,barrett_accuracy_2018,bray_neutron_2018,giacobbo_progenitors_2018,kruckow_progenitors_2018,taylor_mining_2018,vigna-gomez_formation_2018}.
As our understanding of uncertain binary-evolution processes advances, updated models can be implemented into BPS codes.
\par
We have previously shown that SN mass loss in close binaries can produce GWs that are larger in magnitude than those produced by orbital motion \citep{holgado_gravitational_2019}. 
Here we use this analysis to model SN mass loss and SN natal kicks in close binaries containing helium stars with NS companions.
From the dynamical evolution of the binaries, we compute the expected GW emission, estimate the back-reaction on the post-SN orbit, and estimate prospects for detectability by next-generation mid-band GW detectors such as DECIGO \citep[e.g.,][]{sato_decigo:_2009,sato_status_2017}.
\par
We focus on systems with NSs since there exist physical constraints on NS kicks from Galactic double NS observations \citep[e.g.,][]{iben_origin_1996,willems_pulsar_2004,hobbs_statistical_2005,wang_neutron_2006,wong_constraints_2010,andrews_evolutionary_2015,andrews_double_2019,andrews_double_2019-1} and Galactic SN remnants \citep[e.g.,][]{holland-ashford_comparing_2017,katsuda_intermediate-mass_2018,schinzel_tail_2019}.
BH natal kicks are not as well constrained; however, some indirect evidence exists for slow BH natal kicks from observations of low-mass X-ray binaries \citep[e.g.,][]{repetto_investigating_2012,mandel_estimates_2016,atri_potential_2019}.
Conversely, the detection of GW151226 \citep{ligo_scientific_collaboration_and_virgo_collaboration_gw151226:_2016} and constraints on its spin-misalignment have provided indirect evidence for fast BH natal kicks \citep{oshaughnessy_inferences_2017}.
We thus leave BH kicks for future investigation. 
We show that GWs may carry away a significant fraction of the post-SN orbital energy and may thus be an important process to treat in BPS codes. 
\par
This paper is organized as follows. 
In \S\ref{sec:sne}, we review previous treatments of natal kicks in asymmetric SNe that are often used in BPS codes. 
In this section, we also review aspects of GW emission relevant to this work. 
In \S\ref{sec:methods}, we then describe our method for modeling the transition from the pre-SN binary to post-SN binary and estimating the GW signal from this transition.
In \S\ref{sec:randd}, we present our results for the GW energy and angular-momentum distributions and estimate prospects for detectability with next-generation mid-band GW detectors like DECIGO.
In \S\ref{sec:con}, we summarize our findings and discuss future work.
\section{Supernova mass loss, natal kicks, and gravitational radiation} \label{sec:sne}
%The coordinate system we consider is illustrated in Figure \ref{fig:coord}.
After a NS emerges from the common-envelope phase \citep[e.g.,][]{macleod_accretion-fed_2014,holgado_gravitational_2018,fragos_complete_2019}, a remnant helium (He) star with mass $m_{\rm He}$ and a primary NS with mass $m_1$ are expected to comprise the post-common-envelope binary. 
The He star is then expected to eventually undergo a SN explosion, which may either form a BNS that merges within a Hubble time, a BNS that is effectively stalled, an unbound BNS, or a NSBH system. 
The recent detection of a candidate ultra-stripped SN \citep{de_hot_2018} has provided additional strong evidence that HeNS binaries are the progenitors of BNSs \citep[e.g.,][]{dewi_evolution_2002,ivanova_role_2003,tauris_ultra-stripped_2013,tauris_ultra-stripped_2015}.
\subsection{The instantaneous limit}
We take pre-SN binaries to be circular for this work and have an initial semi-major axis $a_{\rm i}$ and total mass $M_{\rm i} = m_1 + m_{\rm He}$, with an initial orbital period that obeys Kepler's Third Law: $P_{\rm i}^2 = 4 \pi^2 a_{\rm i}^3 / (GM)$. 
The energy and angular momentum of the circular binary are
\begin{align}
E_{\rm i} &= - \frac{G \mu_{\rm i} M_{\rm i}}{2a_{\rm i}} \ , \\
|{\bf J}_{{\rm i}} |&= \mu_{\rm i} \sqrt{G M_{\rm i} a_{\rm i}} \ ,
\end{align}
where $\mu_{\rm i} = m_1 m_{\rm He}/M_{\rm i}$ is the initial reduced mass.
BPS codes assume that the transition from the pre-SN binary to the post-SN binary is effectively instantaneous and that the energy and specific angular momentum are conserved. 
Under these assumptions, the post-SN semi-major axis $(a_{\rm f})$ and eccentricity $(e_{\rm f})$ of a kicked bound binary obey \citep[e.g.,][]{kalogera_orbital_1996,postnov_evolution_2006}
\begin{subequations} \label{eq:bps}
\begin{align} \label{eq:abps}
a_{\rm f} &= G \left(m_1 + m_2\right) \left(\frac{2 G (m_1 + m_2)}{a_{\rm i}} - {\rm v}_{\rm k}^2 - {\rm v}_{\rm r}^2 - 2 {\rm v}_{{\rm k},y} {\rm v}_{\rm r}\right)^{-1} , \\ \label{eq:ebps}
1 &- e_{\rm f}^2 = \frac{a_{\rm i}^2 \left({\rm v}_{{\rm k},z}^2 + {\rm v}_{{\rm k},y}^2 + {\rm v}_{\rm r}^2 + 2 {\rm v}_{{\rm k},y} {\rm v}_{\rm r}\right)}{G (m_1 + m_2) a_{\rm f}} \ ,
\end{align}
\end{subequations}
where $m_2$ is the secondary NS mass, ${\rm v}_{\rm r}$ is the magnitude of the relative orbital velocity between the primary and secondary, and ${\rm v}_{\rm k}$ is the magnitude of the kick velocity.
In \autoref{eq:bps}, the coordinate system is oriented such that ${\bf v}_{\rm r}$ is parallel to the $y$ axis and the orbital plane coincides with the $xy$ plane. 
\subsection{Gravitational radiation}
The mass loss and natal kick from an asymmetric SN explosion accelerate the nascent compact object, contributing to the binary's gravitational radiation.
We treat both the primary and secondary as point masses such that the mass-quadrupole tensor is $I_{jk} = \mu r_j r_k$, where $\mu = \mu(t)$ is the reduced mass and ${\bf r} = (x,y,z)$ is the relative separation in Cartesian coordinates.
The distribution of ejecta following the SN explosion may torque the remnant binary and also contribute to the quadrupole moment of the system as a whole, though we assume that these are negligible for this work in order to make consistent comparisons with what BPS codes model, i.e., \autoref{eq:bps}.  
\par
We have previously shown \citep{holgado_gravitational_2019} that the third time derivatives of the mass-quadrupole tensor components in Cartesian coordinates are 
\begin{subequations}
\begin{align}
\dddot{I}_{xx} &= \dddot{\mu} x^2 + 6 \ddot{\mu} x \dot{x} + 6 \dot{\mu} \dot{x}^2 + 6 \dot{\mu} x \ddot{x} + 6 \mu \dot{x} \ddot{x} + 2 \mu x \dddot{x} \ , \\
\dddot{I}_{yy} &= \dddot{\mu} y^2 + 6 \ddot{\mu} y \dot{y} + 6 \dot{\mu} \dot{y}^2 + 6 \dot{\mu} y \ddot{y} + 6 \mu \dot{y} \ddot{y} + 2 \mu y \dddot{y} \ , \\
\dddot{I}_{zz} &= \dddot{\mu} z^2 + 6 \ddot{\mu} z \dot{z} + 6 \dot{\mu} \dot{z}^2 + 6 \dot{\mu} z \ddot{z} + 6 \mu \dot{z} \ddot{z} + 2 \mu z \dddot{z} \ , \\
\dddot{I}_{xy} &= \dddot{I}_{yx} = \dddot{\mu} xy + 3 \ddot{\mu} \dot{x} y + 3 \dot{\mu} \ddot{x} y + \mu \dddot{x} y + 3 \ddot{\mu} x \dot{y} \\ \nonumber &\phantom{yx = =} + 6 \dot{\mu} \dot{x} \dot{y} + 3 \mu \ddot{x} \dot{y} + 3 \dot{\mu} x \ddot{y} + 3 \mu \dot{x} \ddot{y} + \mu x \dddot{y} \ , \\
\dddot{I}_{xz} &= \dddot{I}_{zx} = \dddot{\mu} xz + 3 \ddot{\mu} \dot{x} z + 3 \dot{\mu} \ddot{x} z + \mu \dddot{x} z + 3 \ddot{\mu} x \dot{z} \\ \nonumber &\phantom{zx = =} + 6 \dot{\mu} \dot{x} \dot{z} + 3 \mu \ddot{x} \dot{z} + 3 \dot{\mu} x \ddot{z} + 3 \mu \dot{x} \ddot{z} + \mu x \dddot{z} \ , \\
\dddot{I}_{yz} &= \dddot{I}_{zy} = \dddot{\mu} yz + 3 \ddot{\mu} \dot{y} z + 3 \dot{\mu} \ddot{y} z + \mu \dddot{y} z + 3 \ddot{\mu} y \dot{z} \\ \nonumber &\phantom{zy = =} + 6 \dot{\mu} \dot{y} \dot{z} + 3 \mu \ddot{y} \dot{z} + 3 \dot{\mu} y \ddot{z} + 3 \mu \dot{y} \ddot{z} + \mu y \dddot{z} \ ,
\end{align}
\end{subequations}
which we use here to estimate the GW energy and GW angular momentum radiated away during the binary phase transition. 
GWs will also be produced by the SN explosion itself in the LIGO band due to the convective fluid motions at core collapse and subsequent proto-NS oscillations \citep[e.g.,][]{fryer_gravitational_2003,fryer_gravitational_2004,gossan_observing_2016,powell_inferring_2016,andresen_gravitational_2017,morozova_gravitational_2018,andresen_gravitational_2019,pajkos_features_2019,powell_gravitational_2019,radice_characterizing_2019}, but we do not focus on these for this work. 
%The GW strain of the binary is
%
%\begin{equation}
%h_{jk} = \frac{2G}{c^4 D} \ddot{\cal I}_{jk} \ ,
%\end{equation}
%
%where $D$ is the distance to the source, and where we take the plus and cross polarizations, $h_+$ and $h_\times$, to be projected along the line of sight.
%We will assume that the ejecta are at low-enough densities such that their contribution to the mass quadrupole moment. 
%
\section{Methods} \label{sec:methods}
\subsection{Governing equations}
We model the transition from the pre-SN binary to the post-SN binary with the following equation of motion (EOM)
\begin{equation} \label{eq:eom}
\ddot{r}_j = - \frac{GM}{r^3} r_j + \dot{\rm v}_{{\rm k},j} \ ,
\end{equation}
where $\dot{\rm v}_{{\rm k},j}$ is the $j$th component of the kick acceleration and where we have assumed that the hydrodynamical interactions during the SN phase transition \citep[e.g.,][]{wheeler_supernovae_1975,fryxell_hydrodynamic_1981} to be negligible. 
In order to ensure that the time derivatives of the quadrupole moment are well defined for all $t$, we take the mass loss and kick acceleration to smoothly vary with time via
\begin{equation}
\frac{\dot m}{\Delta m} = \frac{\dot{\rm v}_{{\rm k},j}}{{\rm v}_{{\rm k},j}} = - \frac{\exp[(t-t_0)/\tau]}{\tau \left(\exp[(t-t_0)/\tau]+1\right)^2}\ ,
\end{equation}
where $\dot{m}$ is the mass variation rate of the secondary, $\Delta m = m_{\rm He} - m_2$ is the mass lost during the SN explosion, $t_0$ is the time at which the peak of the mass loss and natal kick occurs, and $\tau$ is the characteristic timescale for the mass loss $(\tau_{\rm m})$ and natal kick $(\tau_{\rm k})$.
The explicit form for the mass of the secondary is $m(t) = m_2 + \Delta m \left(\exp[(t-t_0)/\tau_{\rm m}] + 1\right)^{-1}$.
For this work, we take $t_0 = P_{\rm i}$. 
For bound post-SN binaries, we estimate the post-SN semi-major axis $(a_{\rm f,EOM})$ and eccentricity $(e_{\rm f,EOM})$ by computing the periapsis $(r_{\rm p})$ and apoapsis $(r_{\rm a})$ distances using 
\begin{subequations} \label{eq:reom}
\begin{align}
%\begin{equation} \label{eq:reom}
r_{\rm p} &= \min (r) = a_{\rm f,EOM} \left(1-e_{\rm f,EOM}\right) \ , \\
r_{\rm a} &= \max (r) = a_{\rm f,EOM} \left(1+e_{\rm f,EOM} \right) \ ,
%\end{equation}
\end{align}
\end{subequations}
and solving both expressions for the post-SN semi-major axis $a_{\rm f,EOM}$ and eccentricity $e_{\rm f,EOM}$.
\par
\begin{figure}
\centering
\includegraphics[width=\columnwidth]{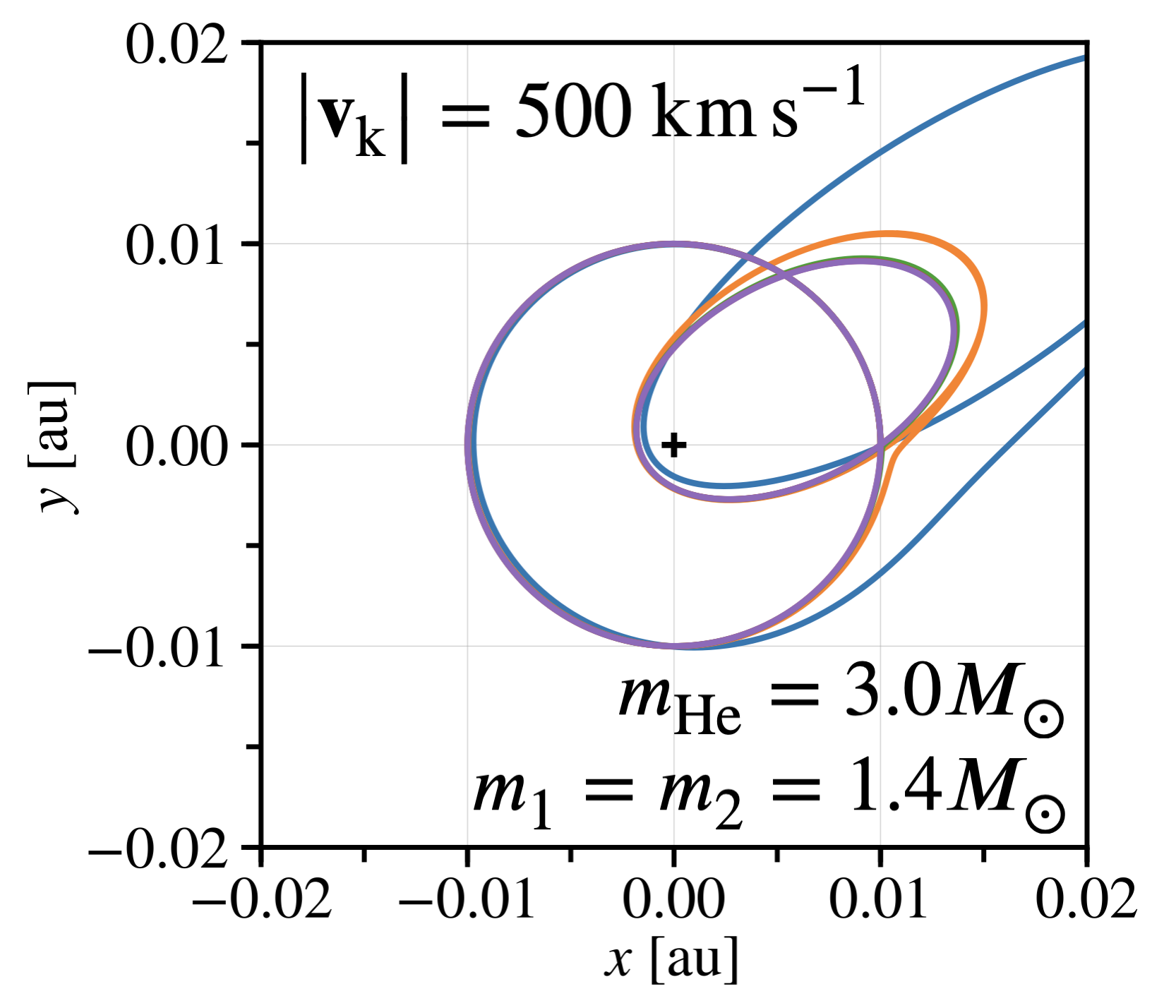}
\includegraphics[width=\columnwidth]{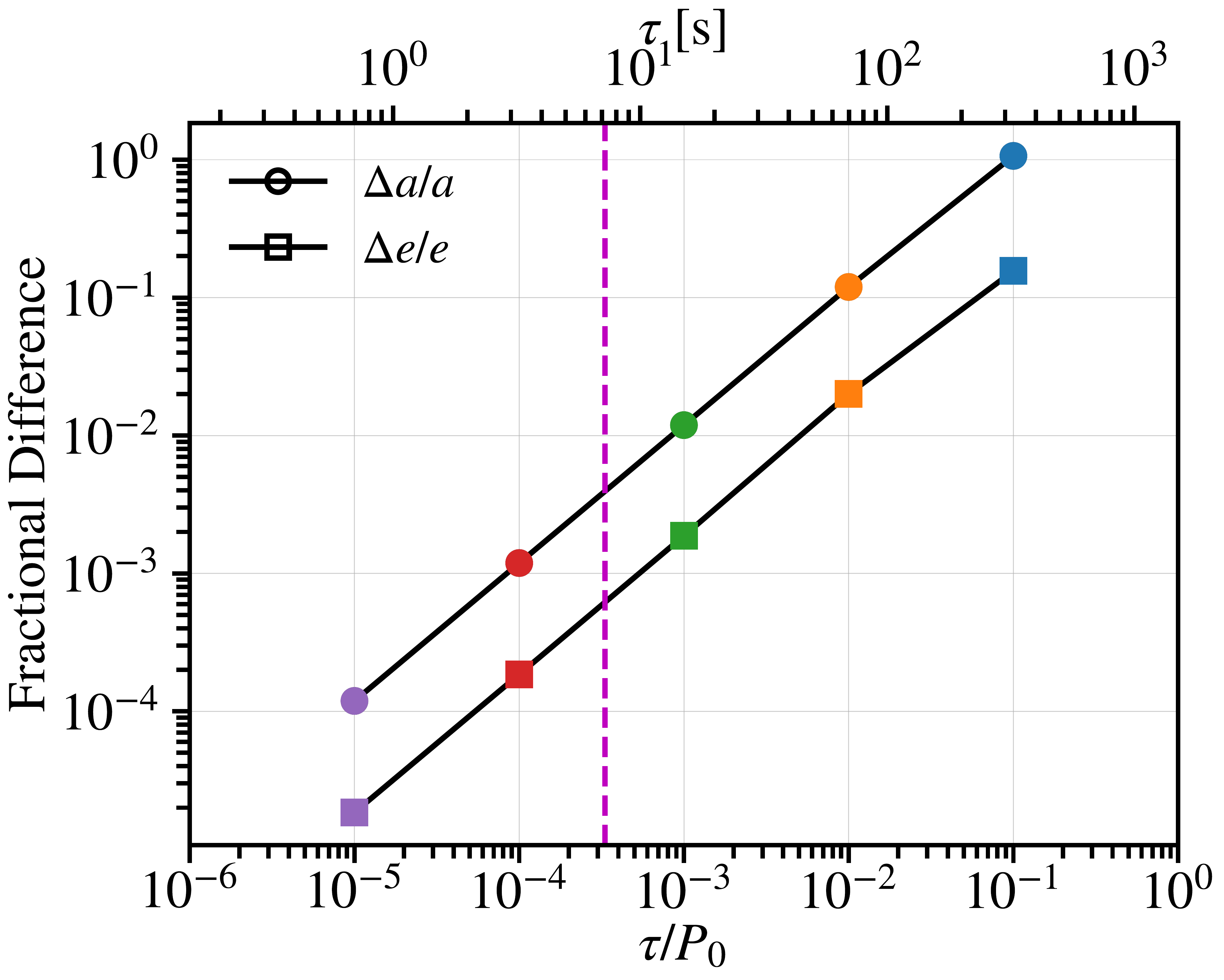}
\caption{\label{fig:converge} Top panel: orbital trajectories for $\tau /P_{\rm i} = \tau_{\rm m}/P_{\rm i} = \tau_{\rm k}/P_{\rm i} = \left(10^{-5},10^{-4},10^{-3},10^{-2},10^{-1}\right)$.
Bottom panel: fractional difference of the predicted semi-major axis $\Delta a/a = |a_{\rm f} - a_{\rm f,EOM}|/a_{\rm f}$ (line with circle markers) and eccentricity $\Delta e/e = |e_{\rm f} - e_{\rm f,EOM}|/e_{\rm f}$ (line with square markers) of the post-SN binary between our model and BPS. 
The color of each filled circle and filled square corresponds to the color of the orbital trajectory in the top panel.
The dashed magenta line is the light-crossing time for the initial circular binary.
}
\end{figure}

As a consistency check for this model, we perform a convergence test for Equations~\ref{eq:eom}-\ref{eq:reom} to compare with \autoref{eq:bps}. 
For our convergence test, we take $m_{\rm He} = 3.0 M_\odot$, $m_1 = m_2 = 1.4 M_\odot$, ${\bf v}_{\rm k} = 500 \left(1/\sqrt{2},-1/\sqrt{2},0\right) \, {\rm km\ s}^{-1}$, and $a_{\rm i} = 10^{-2} \, {\rm au}$, and we choose a set of mass-loss and kick timescales $\tau/P_{\rm i} = \tau_{\rm m}/P_{\rm i} = \tau_{\rm k}/P_{\rm i} = \left(10^{-5}, 10^{-4}, 10^{-3}, 10^{-2}, 10^{-1}\right)$. 
In \autoref{fig:converge}, we plot the orbital trajectories in the top panel for the different timescales we consider.
In the bottom panel of \autoref{fig:converge}, we plot the fractional differences of the predicted post-SN semi-major axis ($\Delta a/a = |a_{\rm f} - a_{\rm f,EOM}|/a_{\rm f}$) and eccentricity ($\Delta e/e = |e_{\rm f} - e_{\rm f,EOM}|/e_{\rm f}$) between the two models.
The fractional differences of the post-SN semi-major axis and eccentricity between the two models decreases as the timescales decrease, verifying that our model converges to BPS in the limit of instantaneous mass loss and natal kicks.
\subsection{GW bursts}
The GW signals produced by the SN mass loss and natal kick are bursts: short-duration peaks in the strain. 
Here, we estimate the GW energy $(E_{\rm GW})$ and the GW angular momentum $(J_{{\rm GW},i})$ radiated during a burst to leading order using
\begin{subequations}
\begin{align}
E_{\rm GW} &\approx \frac{1}{5} \frac{G}{c^5}\int_{\Delta t} \dddot{I}_{jk} \dddot{I}_{jk} \ {\rm d}t \ , \\
J_{{\rm GW}, i} &\approx \frac{2}{5} \frac{G}{c^5} \int_{\Delta t} \epsilon_{ijk} \ddot{I}_{jm} \dddot{I}_{km} \ {\rm d}t \ ,
\end{align}
\end{subequations}
where $\Delta t$ is a chosen time interval centered at the peak of the mass loss and natal kick and $\epsilon_{ijk}$ is the Levi-Civita tensor. 
We emphasize that this is a leading-order estimate of the GW energy and angular momentum radiated from the SN mass loss and natal kick.
Since the kick timescale is much shorter than the mass-loss timescale, we expect that GW emission from the natal kick will be much larger than that from the mass loss. 
\par
The observed strain on Earth depends on the orientation of the binary relative to the line of sight. 
The binary can be randomly oriented with respect to the line of sight, and the kick direction can be randomly oriented with respect to the orbital plane. 
In the transverse-traceless gauge, the observed strain tensor is 
\begin{equation}
h_{jk}^{\rm TT} = \frac{2G}{c^4 D} \ddot{\cal I}_{jk}^{\rm TT} = \frac{2G}{c^4 D} \left( {\cal P}_{jl} {\cal P}_{km} \ddot{\cal I}_{l m} - \frac{1}{2} {\cal P}_{jk} {\cal P}_{lm} \ddot{\cal I}_{lm}\right) \ ,
\end{equation}
where ${\cal I}_{jk} = I_{jk} - \frac{1}{3} \delta_{jk} I_{jk}$ is the reduced quadrupole moment tensor, $D$ is the luminosity distance to the source, and ${\cal P}_{jk} = \delta_{jk} - n_j n_k$ is the projection tensor from the orbital reference frame to the observer's frame. We take the unit vector along the line of sight in the observer's frame to be ${\bf n} = (0,0,1)$. 
With this line-of-sight orientation, the strain plus and cross polarizations are $h_+ = h_{xx} = - h_{yy}$ and $h_\times = h_{xy} = h_{yx}$, respectively (where $x$ and $y$ here refer to the observer's frame). 
%The root-sum-square strain of the GW burst is
%
%\begin{equation}
%h_{\rm rss} = \int \left(\left|h_+\right|^2 + \left|h_\times\right|^2\right) \, {\rm d}t \ .
%\end{equation}
%
\par
The signal-to-noise ratio (SNR) of the GW burst (prior to modulation by the detector antenna pattern) obeys
\begin{equation} \label{eq:snr}
{\rm SNR} = \sqrt{4 \int_{-\infty}^\infty \frac{|\tilde{h}(f)|^2}{{\cal S}(f)} \, {\rm d}f} \ ,
\end{equation}
where $f$ is the GW frequency, $\tilde{h}(f)$ is the Fourier-transformed strain and ${\cal S}(f)$ is the detector strain sensitivity. 
For this work, we ignore Doppler boosting of the observed GW strain and GW frequency due to the motion of the binary relative to the line of sight since the largest natal kick velocities we consider satisfy ${\rm v}_{\rm k}/c \lesssim 1\%$. 
\par
A variety of methods have been developed for GW burst detection \citep[e.g.,][]{anderson_excess_2001,pradier_efficient_2001}. 
Here, we are primarily interested in determining the distances at which GW bursts from natal kicks will produce statistically significant excess power in a detector. 
Often a fiducial SNR threshold of 4.5 for GW burst detectability is used \cite[e.g.,][]{abbott_prospects_2016,macleod_fully-coherent_2016}, and we adopt this criterion here.
\section{Results and Discussion} \label{sec:randd}
\subsection{Energy and angular momentum in GWs}
We first investigate the GW energy and angular momentum distributions for both bound and unbound post-SN binaries, using the mass constraints on GW170817 as a fiducial system.  
In \autoref{tab:params}, we tabulate the model parameter ranges that we consider. 
We compute $5{\times}10^4$ models for bound binaries and $5{\times}10^4$ models for unbound binaries (a total of $10^5$ models) with initial parameters sampled from the tabulated distributions. 
\begin{table} %[b!]
%\hspace*{-1cm}
\centering
%\resizebox{\columnwidth}{!}{%
\begin{tabular}{c l l}
\hline
Parameter & Description & Method \\
\hline\hline
$m_{1}$ & Primary NS mass & Uniform $[1.36 M_\odot, 1.60 M_\odot]$ \\
$m_{2}$ & Secondary NS mass  & Uniform $[1.17 M_\odot, 1.36 M_\odot]$ \\
$m_{\rm He}$ & Helium-star mass & Uniform $[2.6 M_\odot, 8.0 M_\odot]$ \\
$a_{\rm i}$ & Pre-SN semi-major axis & Log uniform $[0.1 R_\odot, 10R_\odot]$ \\
${\rm v}_{\rm k}$ & SN kick velocity & Uniform $[0.0, 2500 ] \ {\rm km} \, {\rm s}^{-1}$ \\
$\tau_{\rm m}$ & Mass-loss timescale & Log uniform $[a_{\rm i}/c,10^{-2} P_{\rm i}]$ \\
$\tau_{\rm k}$ & Kick timescale & Log uniform $[0.05,0.5] \ {\rm s}$\\
$a_{\rm f}$ & BPS post-SN semi-major axis & \autoref{eq:abps} \\
$e_{\rm f}$ & BPS post-SN eccentricity & \autoref{eq:ebps} \\
$a_{\rm f,EOM}$ & EOM post-SN semi-major axis & Equations~\ref{eq:eom}-\ref{eq:reom} \\
$e_{\rm f,EOM}$ & EOM post-SN eccentricity & Equations~\ref{eq:eom}-\ref{eq:reom} \\
\hline
\end{tabular}%
%}
\caption{\label{tab:params} Table of model parameters with descriptions and methods with which they are sampled or calculated. 
}
\end{table}
The NS masses are uniformly distributed with $m_1 \in [1.36 M_\odot, 1.60 M_\odot]$ and $m_2 \in [1.17 M_\odot, 1.36 M_\odot]$.
The mass-loss and kick timescales are sampled log-uniformly in $\left[a_{\rm i}/c, 10^{-2} P_{\rm i}\right]$ and $\left[0.05, 0.5\right] \, {\rm s}$, respectively. 
The mass-loss timescale is bounded by the light-crossing time of the binary, and the kick timescales are informed by multi-dimensional core-collapse SN simulations \citep[e.g.,][]{gessner_hydrodynamical_2018}.
The magnitude of the kick velocities is sampled uniformly in $\left[0, 2500\right] \, {\rm km\ s}^{-1}$, with the kick direction sampled isotropically on the unit sphere. 
\par
\begin{figure}
\centering
\includegraphics[width=\columnwidth]{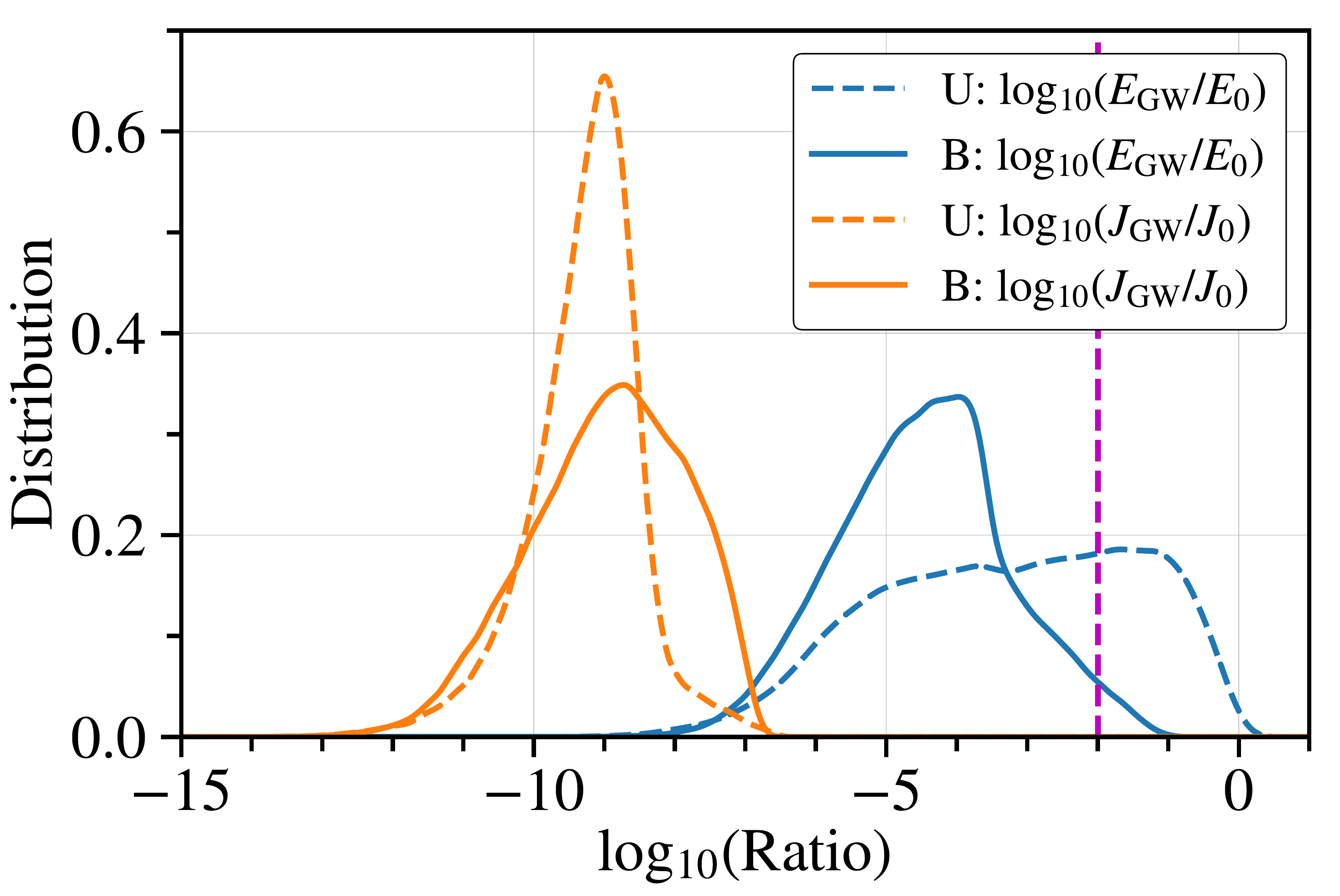}
\includegraphics[width=\columnwidth]{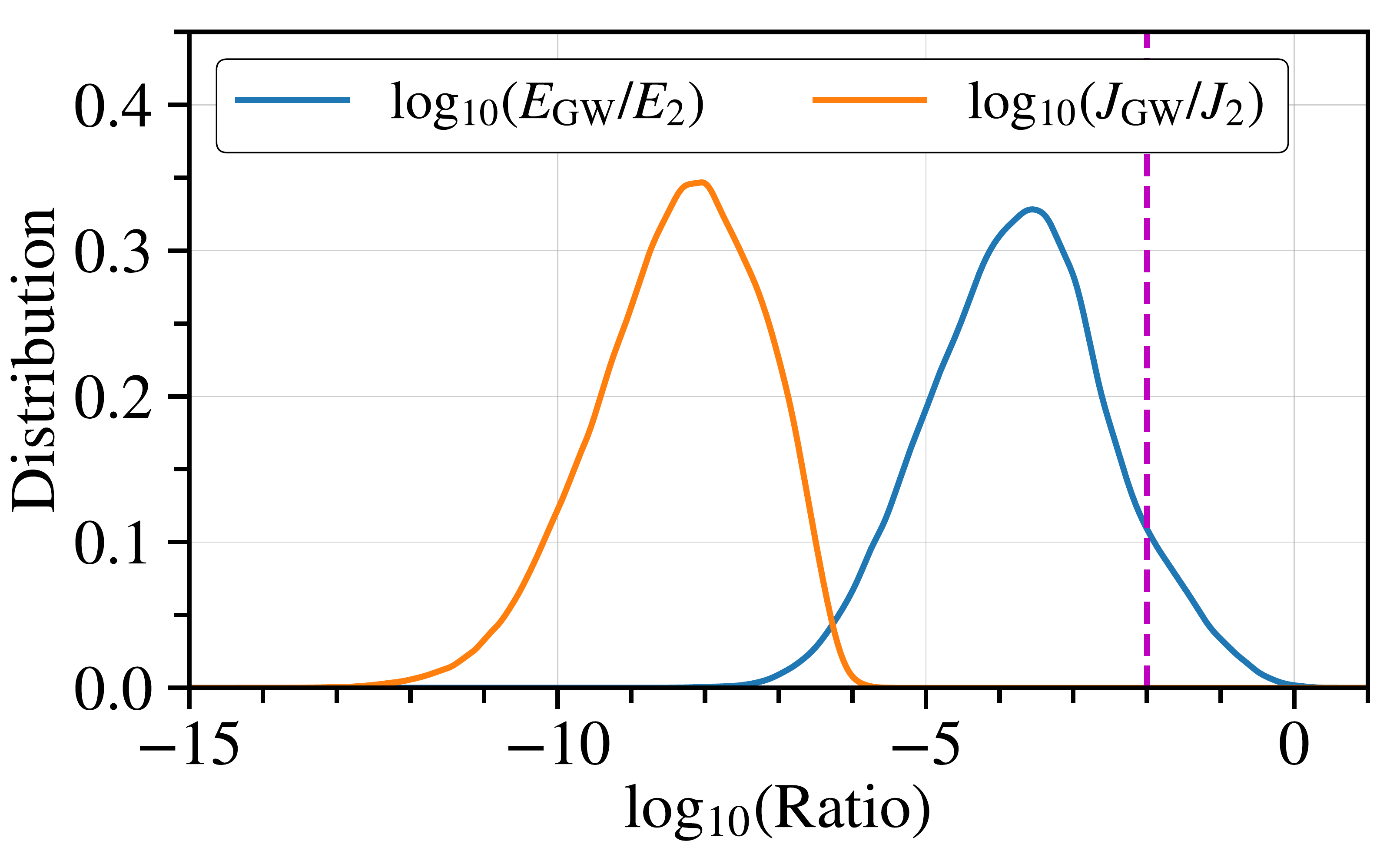}
\caption{\label{fig:dist} Top panel: distribution of energy and angular momentum radiated in GWs for $5{\times}10^4$ bound (B) and $5{\times}10^4$ unbound (U) binaries relative to their initial energy and angular momenta.
The blue and orange filled lines correspond to the energy and angular-momentum ratios for bound binaries, respectively. 
The blue and orange dashed lines correspond to the same quantities for unbound binaries. 
The dashed magenta line corresponds to a value of $1\%$. 
Bottom panel: distribution of energy and angular momentum radiated in GWs for bound binaries relative to the post-SN binary orbital energy and angular momentum. 
}
\end{figure}

We plot comparisons of GW energy and angular momentum between bound and unbound binaries relative to their initial energy and angular momenta in the top panel of \autoref{fig:dist}.
%
%We plot the distribution of semi-major axis. 
% ``
%\begin{figure}
%\centering
%\includegraphics[width=\columnwidth]{da.pdf}
%\caption{\label{fig:da} Top panel: distribution of energy radiated in GWs for bound and unbound binaries. 
%}
%\end{figure}
%
The unbound binaries have a larger fraction of GW energy emitted above $1\%$ compared to the bound binaries. 
This is because larger kick velocities and shorter kick timescales not only are more likely to unbind the binary, they are also more likely to produce larger-amplitude GW bursts. 
We find that the angular momentum radiated away in GWs is a negligible fraction of the initial binary angular momentum for bound and unbound binaries. 
\par
We also plot the the distributions of GW energy and angular momentum relative to those of the post-SN binary. 
A fraction of bound binaries has an estimated GW energy $\gtrsim 10\%$ of the post-SN binary, so their semi-major axes will be comparably smaller than what BPS would predict, since the latter assumes energy conservation. 
As for the pre-SN comparison, the angular momentum radiated in GWs is a negligible fraction of the post-SN binary's angular momentum. 
\subsection{Detectability}
\begin{figure}
\centering
\includegraphics[width=\columnwidth]{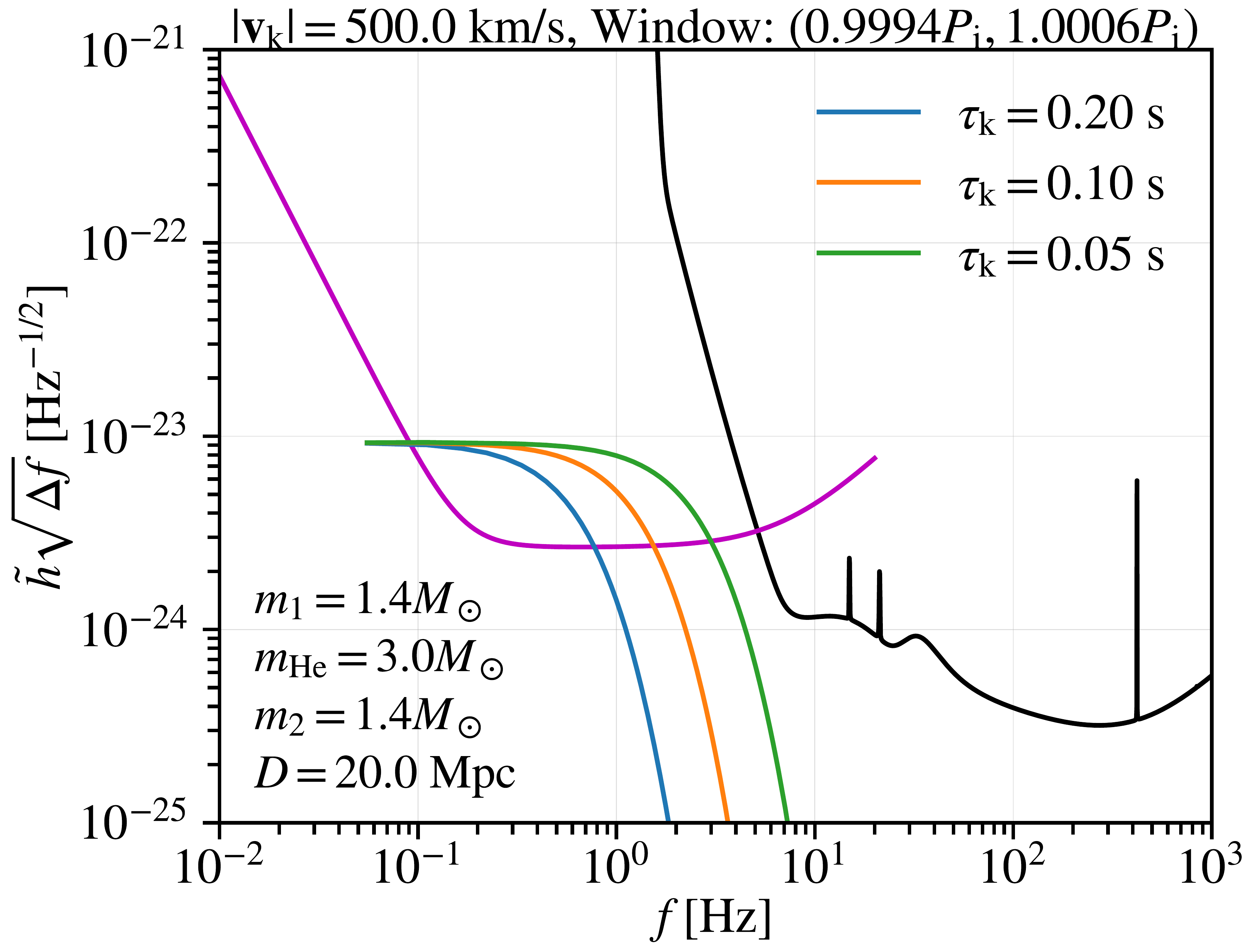}
\includegraphics[width=\columnwidth]{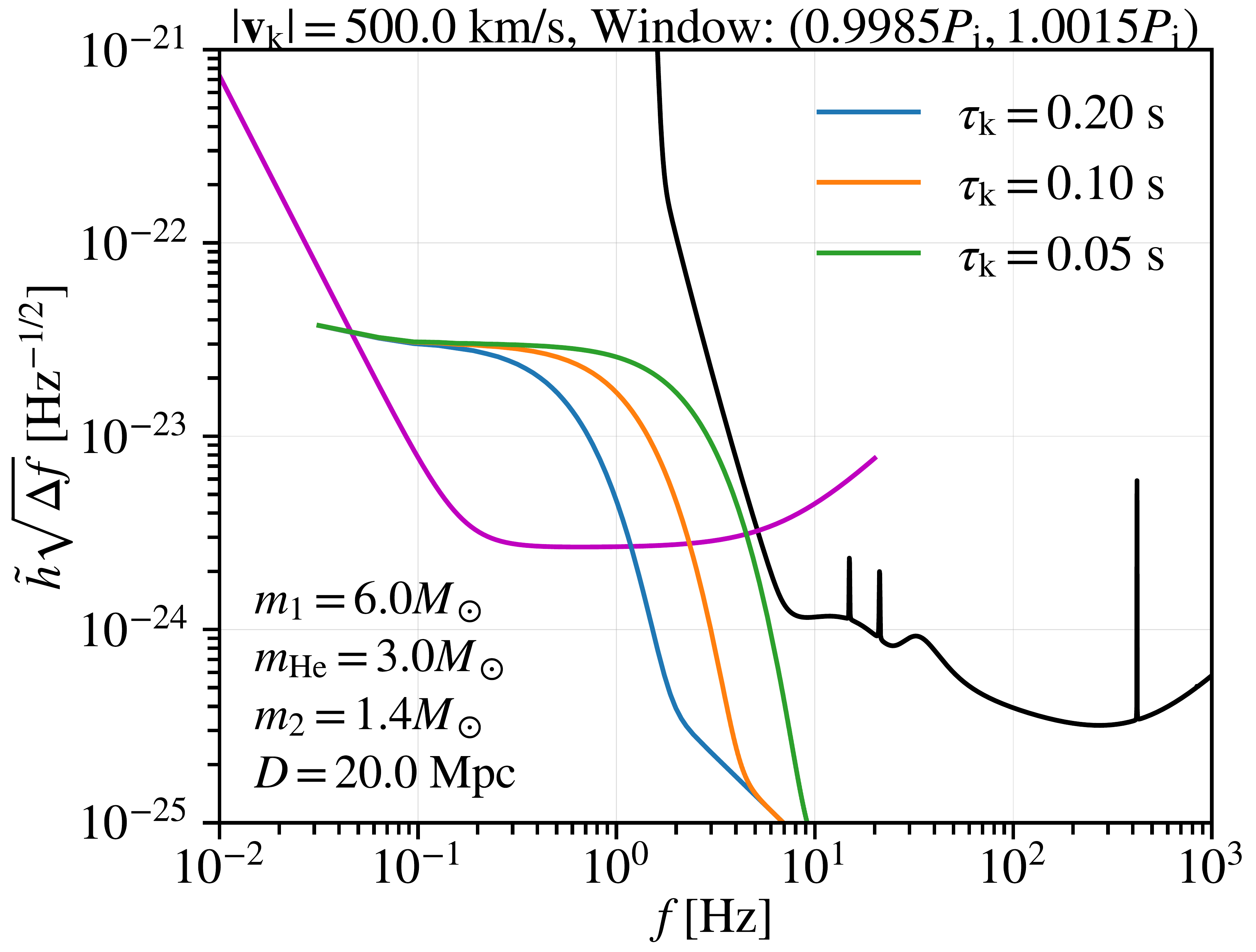}
\caption{\label{fig:burst} Top panel: Fourier transform of the strain from a kicked binary with $m_1 = m_2 = 1.4 M_\odot$, corresponding to a BNS. 
The blue, orange, and green colors correspond to kick timescales of $\tau_{\rm k} = 0.20 \, {\rm s}$, $0.10 \, {\rm s}$, and $0.05 \, {\rm s}$, respectively. 
The window size for this case is $(0.9994P_{\rm i}, 1.0006 P_{\rm i})$. 
The magenta and black lines are the sensitivity curves for DECIGO and the Einstein Telescope, respectively. 
Bottom panel: same as the top panel, but for a binary with $m_1 = 6.0 M_\odot$, corresponding to a NSBH binary. 
The window size for this case is $(0.9985P_{\rm i}, 1.0015 P_{\rm i})$. 
}
\end{figure}

We plot the Fourier-transformed strain for a binary at $D = 20 \, {\rm Mpc}$ in the top panel of \autoref{fig:burst} along with the strain sensitivity curves for DECIGO \citep[e.g.,][]{yagi_detector_2011} and the Einstein Telescope \citep[e.g.,][]{hild_sensitivity_2011}.
The Einstein Telescope is designed to be more sensitive to lower frequencies compared to Cosmic Explorer \citep{abbott_exploring_2017}. 
For shorter natal-kick timescales, the strain spectrum extends to higher GW frequencies, increasing the area between the strain spectrum and the DECIGO sensitivity curve and thus increasing the SNR. 
\par
In the bottom panel of \autoref{fig:burst}, we plot Fourier spectra for a NSBH binary with a BH mass of $m_1 = 6.0 M_\odot$. 
As expected, natal-kick GW bursts from NSBH binaries will be detectable to larger distances compared to BNSs due to the higher chirp mass.
The event rate for NSBH mergers, however, is expected to be lower than that of BNSs \citep[e.g.,][]{mink_merger_2015,gupta_implications_2017,mapelli_cosmic_2018}, which implies that the SN event rate for NSBH binaries will also be lower.
BBH natal-kick GW bursts would also be detectable to larger distances if BHs receive SN natal kicks with strengths and timescales comparable to NSs. 
\par
\begin{figure}
\centering
\includegraphics[width=\columnwidth]{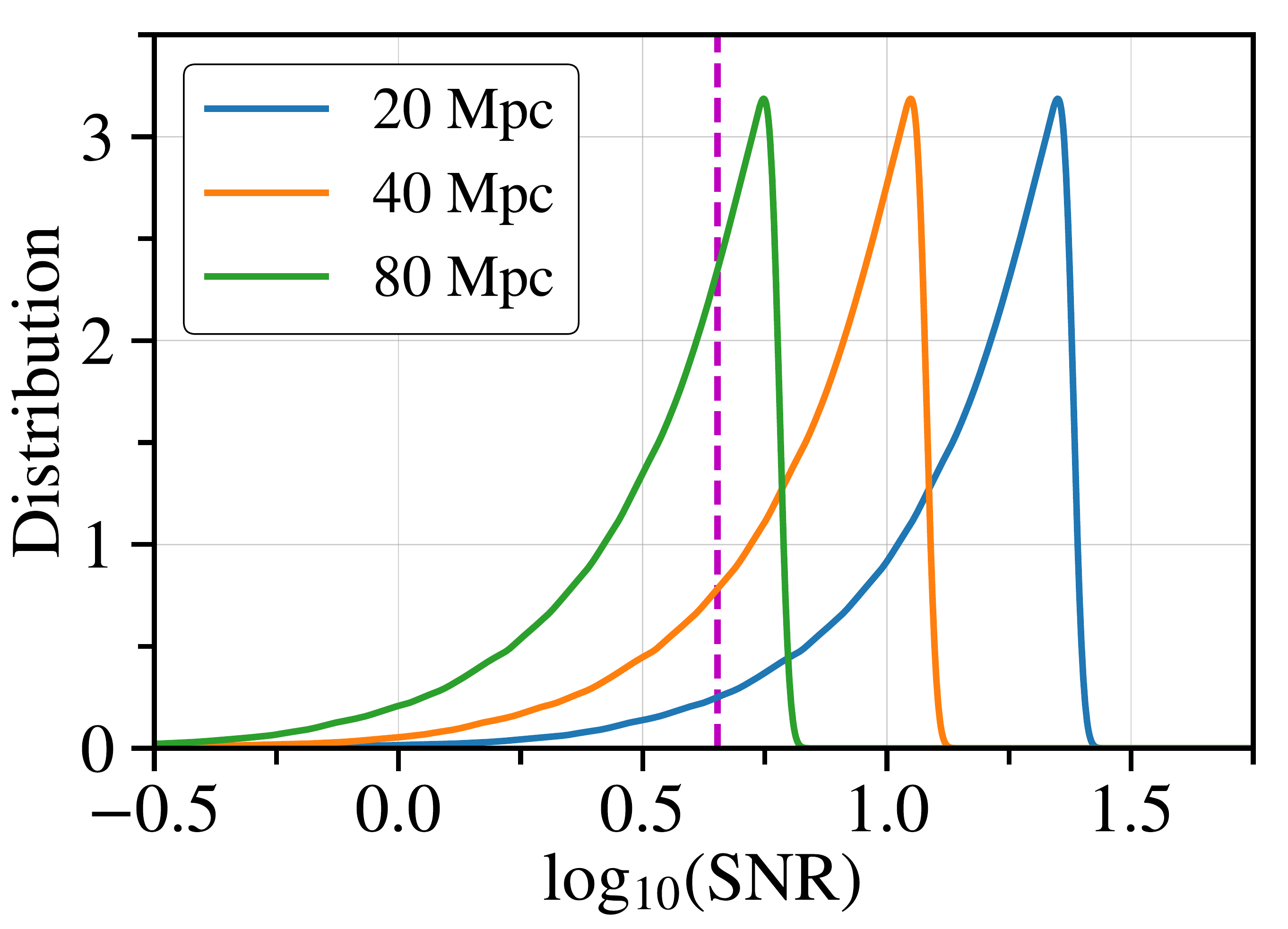}
\caption{\label{fig:snrs} Distributions of the SNR (as calculated via \autoref{eq:snr}) for binary parameters of $m_1 = m_2 = 1.4 M_\odot$, $m_{\rm He} = 3.0 M_\odot$, $a_{\rm i} = 10^{-2} \ {\rm au}$, $|{\bf v}_{\rm k}| = 500 \ {\rm km\ s}^{-1}$, and $\tau_{\rm k} = 0.1 \, {\rm s}$.
We sample the kick directions and orbital-plane orientations relative to the line-of-sight isotropically, with $10^3$ kick directions and for each kick direction, $10^3$ line-of-sight orientations ($10^6$ total binary configurations).
The blue, orange, and green colors correspond to the SNR being evaluated at sources distances of $D = 20 \, {\rm Mpc}, 40 \, {\rm Mpc}$, and $80 \, {\rm Mpc}$, respectively. 
The dashed magenta line corresponds to a fiducial SNR threshold of 4.5 for GW burst detectability \citep[e.g.,][]{abbott_prospects_2016,macleod_fully-coherent_2016}. 
The SNR distributions are left-skewed with $94\%$, $80\%$, and $37\%$ lying above the SNR threshold for $D = 20 \, {\rm Mpc}, 40 \, {\rm Mpc}$, and $80 \, {\rm Mpc}$, respectively. 
}
\end{figure}

We now consider how sensitive the SNR of the GW burst is to the initial orientation relative to the line of sight and the kick direction relative to the orbital plane. 
We sample a total of $10^6$ binary configurations, consisting of $10^3$ kick directions and for each individual kick direction, $10^3$ binary orientations relative to the line of sight.
For each configuration, we calculate the SNR of the burst signal (prior to modulation by the detector antenna pattern) at source distances of $D = 20 \, {\rm Mpc}, 40 \, {\rm Mpc}$, and $80 \, {\rm Mpc}$, and plot the corresponding distributions of the SNR in \autoref{fig:snrs}. 
The SNR distributions are left-skewed with 94\%, 80\%, and 37\% lying above the fiducial detectability SNR threshold of 4.5 for $D = 20 \, {\rm Mpc}, 40 \, {\rm Mpc}$, and $80 \, {\rm Mpc}$, respectively.
\par
SNe in close binaries can also be followed up with electromagnetic observations, showing promise for multi-messenger science with decihertz GW detectors. 
Multi-band GW science may also be possible since the core collapse itself will generate GWs at of order hundreds of hertz to kilohertz frequencies. 
Third-generation ground-based detectors like Einstein Telescope and Cosmic Explorer will be sensitive to GWs from the core collapse out to the Magellanic Clouds \citep[e.g.,][]{roma_astrophysics_2019}. 
We thus expect a helium-star SN explosion with a close NS companion may be a multi-band GW source of interest out to such a distance.  
\section{Conclusions} \label{sec:con}
SN mass loss and natal kicks are important processes in the formation of compact binaries in isolation. 
BNSs are expected to primarily form in the field with the immediate progenitor being a helium star/NS system.
As the helium star undergoes a SN explosion, the mass loss and natal kick will reorient the binary and may form either a BNS that merges within a Hubble time, a binary that is effectively stalled, or an unbound binary. 
The SN mass loss and natal kick contribute to the time-varying quadrupole moment of the binary and generate bursts of GWs corresponding to the different timescales of each process. 
\par
We have modeled the transition from the pre-SN binary to the post-SN binary by integrating the equation of motion (\autoref{eq:eom}).
From our model, we have shown that the SN mass loss and natal kicks that form or unbind BNSs produce GW bursts that may be of interest to next-generation mid-band GW detectors like DECIGO.
We find that the energy radiated in the GW burst may be $\gtrsim 10\%$ of the post-SN orbital energy, while the angular momentum radiated away in GWs is negligible, i.e., $\lesssim 10^{-3}\%$. 
It may thus be important to include GW backreaction on the post-SN binary in BPS codes and to determine how this addition affects the expected populations of low-mass X-ray binaries, Galactic binary pulsars, and the merging compact binaries that LIGO detects. 
\par
For future work, we will investigate natal kicks in the isolated-binary channel for BBHs and extend the analysis we present here to a more self-consistent post-Newtonian model for the equation of motion.
We will also investigate how important the 3D hydrodynamical effects may be during the pre-SN to post-SN binary transition as the ejecta leave the binary.  
\section*{Acknowledgements}
AMH is supported by the DOE NNSA Stewardship Science Graduate Fellowship under grant number DE-NA0003864. 
PMR acknowledges support from the NSF under AST 14-13367.
AMH thanks Nico Yunes, Sarah Vigeland, Sean Couch, Deep Chatterjee, Chris Pankow, Mike Zevin, Kyle Kremer, Eve Chase, Fred Rasio, Vicky Kalogera, Mike Pajkos, and Leslie Looney for helpful comments and fruitful discussions. 
AMH especially thanks Luke Kelley and Mike Zevin for hosting a visit to CIERA at Northwestern and Mike Pajkos for his help in verifying our Fourier-transform method. 
This work made use of the Illinois Campus Cluster, a computing resource that is operated by the Illinois Campus Cluster Program in conjunction with the National Center for Supercomputing Applications. 
%
%%%%%%%%%%%%%%%%%%%%%%%%%%%%%%%%%%%%%%%%%%%%%%%%%%



\bibliographystyle{mnras}
\bibliography{references} % if your bibtex file is called example.bib

\begin{thebibliography}{}
\makeatletter
\relax
\def\mn@urlcharsother{\let\do\@makeother \do\$\do\&\do\#\do\^\do\_\do\%\do\~}
\def\mn@doi{\begingroup\mn@urlcharsother \@ifnextchar [ {\mn@doi@}
  {\mn@doi@[]}}
\def\mn@doi@[#1]#2{\def\@tempa{#1}\ifx\@tempa\@empty \href
  {http://dx.doi.org/#2} {doi:#2}\else \href {http://dx.doi.org/#2} {#1}\fi
  \endgroup}
\def\mn@eprint#1#2{\mn@eprint@#1:#2::\@nil}
\def\mn@eprint@arXiv#1{\href {http://arxiv.org/abs/#1} {{\tt arXiv:#1}}}
\def\mn@eprint@dblp#1{\href {http://dblp.uni-trier.de/rec/bibtex/#1.xml}
  {dblp:#1}}
\def\mn@eprint@#1:#2:#3:#4\@nil{\def\@tempa {#1}\def\@tempb {#2}\def\@tempc
  {#3}\ifx \@tempc \@empty \let \@tempc \@tempb \let \@tempb \@tempa \fi \ifx
  \@tempb \@empty \def\@tempb {arXiv}\fi \@ifundefined
  {mn@eprint@\@tempb}{\@tempb:\@tempc}{\expandafter \expandafter \csname
  mn@eprint@\@tempb\endcsname \expandafter{\@tempc}}}

\bibitem[\protect\citeauthoryear{Abbott et~al.,}{Abbott
  et~al.}{2016}]{abbott_prospects_2016}
Abbott B.~P.,  et~al., 2016, \mn@doi [Living Rev. Relativ.]
  {10.1007/lrr-2016-1}, 19

\bibitem[\protect\citeauthoryear{Anderson, Brady, Creighton  \&
  Flanagan}{Anderson et~al.}{2001}]{anderson_excess_2001}
Anderson W.~G.,  Brady P.~R.,  Creighton J. D.~E.,   Flanagan {\'E}.~{\'E}.,
  2001, \mn@doi [PRD] {10.1103/PhysRevD.63.042003}, 63, 042003

\bibitem[\protect\citeauthoryear{Andresen, M\"{u}ller, M\"{u}ller  \&
  Janka}{Andresen et~al.}{2017}]{andresen_gravitational_2017}
Andresen H.,  M\"{u}ller B.,  M\"{u}ller E.,   Janka H.-T.,  2017, \mn@doi
  [MNRAS] {10.1093/mnras/stx618}, 468, 2032

\bibitem[\protect\citeauthoryear{Andresen, M\"{u}ller, Janka, Summa, Gill  \&
  Zanolin}{Andresen et~al.}{2019}]{andresen_gravitational_2019}
Andresen H.,  M\"{u}ller E.,  Janka H.-T.,  Summa A.,  Gill K.,   Zanolin M.,
  2019, \mn@doi [MNRAS] {10.1093/mnras/stz990}, 486, 2238

\bibitem[\protect\citeauthoryear{Andrews \& Mandel}{Andrews \&
  Mandel}{2019}]{andrews_double_2019-1}
Andrews J.~J.,  Mandel I.,  2019, \mn@doi [ApJL] {10.3847/2041-8213/ab2ed1},
  880, L8

\bibitem[\protect\citeauthoryear{Andrews \& Zezas}{Andrews \&
  Zezas}{2019}]{andrews_double_2019}
Andrews J.~J.,  Zezas A.,  2019, \mn@doi [MNRAS] {10.1093/mnras/stz1066}, 486,
  3213

\bibitem[\protect\citeauthoryear{Andrews, Farr, Kalogera  \& Willems}{Andrews
  et~al.}{2015}]{andrews_evolutionary_2015}
Andrews J.~J.,  Farr W.~M.,  Kalogera V.,   Willems B.,  2015, \mn@doi [ApJ]
  {10.1088/0004-637X/801/1/32}, 801, 32

\bibitem[\protect\citeauthoryear{Atri et~al.,}{Atri
  et~al.}{2019}]{atri_potential_2019}
Atri P.,  et~al., 2019, arXiv:1908.07199 [astro-ph]

\bibitem[\protect\citeauthoryear{Barrett, Gaebel, Neijssel, Vigna-G\'{o}mez,
  Stevenson, Berry, Farr  \& Mandel}{Barrett
  et~al.}{2018}]{barrett_accuracy_2018}
Barrett J.~W.,  Gaebel S.~M.,  Neijssel C.~J.,  Vigna-G\'{o}mez A.,  Stevenson
  S.,  Berry C. P.~L.,  Farr W.~M.,   Mandel I.,  2018, \mn@doi [MNRAS]
  {10.1093/mnras/sty908}, 477, 4685

\bibitem[\protect\citeauthoryear{Belczy\'{n}ski \& Bulik}{Belczy\'{n}ski \&
  Bulik}{1999}]{belczynski_effect_1999}
Belczy\'{n}ski K.,  Bulik T.,  1999, A\&A, 346, 91

\bibitem[\protect\citeauthoryear{Beniamini, Hotokezaka  \& Piran}{Beniamini
  et~al.}{2016}]{beniamini_natal_2016}
Beniamini P.,  Hotokezaka K.,   Piran T.,  2016, \mn@doi [ApJL]
  {10.3847/2041-8205/829/1/L13}, 829, L13

\bibitem[\protect\citeauthoryear{Brandt \& Podsiadlowski}{Brandt \&
  Podsiadlowski}{1995}]{brandt_effects_1995}
Brandt N.,  Podsiadlowski P.,  1995, \mn@doi [MNRAS] {10.1093/mnras/274.2.461},
  274, 461

\bibitem[\protect\citeauthoryear{Bray \& Eldridge}{Bray \&
  Eldridge}{2016}]{bray_neutron_2016}
Bray J.~C.,  Eldridge J.~J.,  2016, \mn@doi [MNRAS] {10.1093/mnras/stw1275},
  461, 3747

\bibitem[\protect\citeauthoryear{Bray \& Eldridge}{Bray \&
  Eldridge}{2018}]{bray_neutron_2018}
Bray J.~C.,  Eldridge J.~J.,  2018, \mn@doi [MNRAS] {10.1093/mnras/sty2230},
  480, 5657

\bibitem[\protect\citeauthoryear{Burrows \& Hayes}{Burrows \&
  Hayes}{1996}]{burrows_pulsar_1996}
Burrows A.,  Hayes J.,  1996, \mn@doi [PRL] {10.1103/PhysRevLett.76.352}, 76,
  352

\bibitem[\protect\citeauthoryear{Colpi \& Wasserman}{Colpi \&
  Wasserman}{2002}]{colpi_formation_2002}
Colpi M.,  Wasserman I.,  2002, \mn@doi [ApJ] {10.1086/344405}, 581, 1271

\bibitem[\protect\citeauthoryear{Coulter et~al.,}{Coulter
  et~al.}{2017}]{coulter_swope_2017}
Coulter D.~A.,  et~al., 2017, \mn@doi [Science] {10.1126/science.aap9811}, 358,
  1556

\bibitem[\protect\citeauthoryear{De et~al.,}{De et~al.}{2018}]{de_hot_2018}
De K.,  et~al., 2018, \mn@doi [Science] {10.1126/science.aas8693}, 362, 201

\bibitem[\protect\citeauthoryear{Dewi, Pols, Savonije  \& van~den Heuvel}{Dewi
  et~al.}{2002}]{dewi_evolution_2002}
Dewi J. D.~M.,  Pols O.~R.,  Savonije G.~J.,   van~den Heuvel E. P.~J.,  2002,
  \mn@doi [MNRAS] {10.1046/j.1365-8711.2002.05257.x}, 331, 1027

\bibitem[\protect\citeauthoryear{Dominik, Belczynski, Fryer, Holz, Berti,
  Bulik, Mandel  \& O'Shaughnessy}{Dominik et~al.}{2012}]{dominik_double_2012}
Dominik M.,  Belczynski K.,  Fryer C.,  Holz D.~E.,  Berti E.,  Bulik T.,
  Mandel I.,   O'Shaughnessy R.,  2012, \mn@doi [ApJ]
  {10.1088/0004-637X/759/1/52}, 759, 52

\bibitem[\protect\citeauthoryear{Fragos et~al.,}{Fragos
  et~al.}{2013}]{fragos_x-ray_2013}
Fragos T.,  et~al., 2013, \mn@doi [ApJ] {10.1088/0004-637X/764/1/41}, 764, 41

\bibitem[\protect\citeauthoryear{Fragos, Andrews, Ramirez-Ruiz, Meynet,
  Kalogera, Taam  \& Zezas}{Fragos et~al.}{2019}]{fragos_complete_2019}
Fragos T.,  Andrews J.~J.,  Ramirez-Ruiz E.,  Meynet G.,  Kalogera V.,  Taam
  R.~E.,   Zezas A.,  2019, arXiv:1907.12573 [astro-ph]

\bibitem[\protect\citeauthoryear{Fryer \& Kalogera}{Fryer \&
  Kalogera}{1997}]{fryer_double_1997}
Fryer C.,  Kalogera V.,  1997, \mn@doi [ApJ] {10.1086/304772}, 489, 244

\bibitem[\protect\citeauthoryear{Fryer \& Kusenko}{Fryer \&
  Kusenko}{2006}]{fryer_effects_2006}
Fryer C.~L.,  Kusenko A.,  2006, \mn@doi [ApJS] {10.1086/500933}, 163, 335

\bibitem[\protect\citeauthoryear{Fryer \& New}{Fryer \&
  New}{2003}]{fryer_gravitational_2003}
Fryer C.~L.,  New K. C.~B.,  2003, \mn@doi [Liv. Rev. Relativ.]
  {10.12942/lrr-2003-2}, 6, 2

\bibitem[\protect\citeauthoryear{Fryer, Holz  \& Hughes}{Fryer
  et~al.}{2004}]{fryer_gravitational_2004}
Fryer C.~L.,  Holz D.~E.,   Hughes S.~A.,  2004, \mn@doi [ApJ]
  {10.1086/421040}, 609, 288

\bibitem[\protect\citeauthoryear{Fryxell \& Arnett}{Fryxell \&
  Arnett}{1981}]{fryxell_hydrodynamic_1981}
Fryxell B.~A.,  Arnett W.~D.,  1981, \mn@doi [ApJ] {10.1086/158664}, 243, 994

\bibitem[\protect\citeauthoryear{Gessner \& Janka}{Gessner \&
  Janka}{2018}]{gessner_hydrodynamical_2018}
Gessner A.,  Janka H.-T.,  2018, \mn@doi [ApJ] {10.3847/1538-4357/aadbae}, 865,
  61

\bibitem[\protect\citeauthoryear{Giacobbo \& Mapelli}{Giacobbo \&
  Mapelli}{2018}]{giacobbo_progenitors_2018}
Giacobbo N.,  Mapelli M.,  2018, \mn@doi [MNRAS] {10.1093/mnras/sty1999}, 480,
  2011

\bibitem[\protect\citeauthoryear{Gossan, Sutton, Stuver, Zanolin, Gill  \&
  Ott}{Gossan et~al.}{2016}]{gossan_observing_2016}
Gossan S.,  Sutton P.,  Stuver A.,  Zanolin M.,  Gill K.,   Ott C.,  2016,
  \mn@doi [PRD] {10.1103/PhysRevD.93.042002}, 93, 042002

\bibitem[\protect\citeauthoryear{Gupta, Arun  \& Sathyaprakash}{Gupta
  et~al.}{2017}]{gupta_implications_2017}
Gupta A.,  Arun K.~G.,   Sathyaprakash B.~S.,  2017, \mn@doi [ApJL]
  {10.3847/2041-8213/aa9271}, 849, L14

\bibitem[\protect\citeauthoryear{Hallinan et~al.,}{Hallinan
  et~al.}{2017}]{hallinan_radio_2017}
Hallinan G.,  et~al., 2017, \mn@doi [Science] {10.1126/science.aap9855}, 358,
  1579

\bibitem[\protect\citeauthoryear{Hild et~al.,}{Hild
  et~al.}{2011}]{hild_sensitivity_2011}
Hild S.,  et~al., 2011, \mn@doi [CQG] {10.1088/0264-9381/28/9/094013}, 28,
  094013

\bibitem[\protect\citeauthoryear{Hills}{Hills}{1983}]{hills_effects_1983}
Hills J.~G.,  1983, \mn@doi [ApJ] {10.1086/160871}, 267, 322

\bibitem[\protect\citeauthoryear{Hobbs, Lorimer, Lyne  \& Kramer}{Hobbs
  et~al.}{2005}]{hobbs_statistical_2005}
Hobbs G.,  Lorimer D.~R.,  Lyne A.~G.,   Kramer M.,  2005, \mn@doi [MNRAS]
  {10.1111/j.1365-2966.2005.09087.x}, 360, 974

\bibitem[\protect\citeauthoryear{Holgado \& Ricker}{Holgado \&
  Ricker}{2019}]{holgado_gravitational_2019}
Holgado A.~M.,  Ricker P.~M.,  2019, \mn@doi [ApJ] {10.3847/1538-4357/ab3293},
  882, 39

\bibitem[\protect\citeauthoryear{Holgado, Ricker  \& Huerta}{Holgado
  et~al.}{2018}]{holgado_gravitational_2018}
Holgado A.~M.,  Ricker P.~M.,   Huerta E.~A.,  2018, \mn@doi [ApJ]
  {10.3847/1538-4357/aab6a9}, 857, 38

\bibitem[\protect\citeauthoryear{Holland-Ashford, Lopez, Auchettl, Temim  \&
  Ramirez-Ruiz}{Holland-Ashford et~al.}{2017}]{holland-ashford_comparing_2017}
Holland-Ashford T.,  Lopez L.~A.,  Auchettl K.,  Temim T.,   Ramirez-Ruiz E.,
  2017, \mn@doi [ApJ] {10.3847/1538-4357/aa7a5c}, 844, 84

\bibitem[\protect\citeauthoryear{Hurley, Tout  \& Pols}{Hurley
  et~al.}{2002}]{hurley_evolution_2002}
Hurley J.~R.,  Tout C.~A.,   Pols O.~R.,  2002, \mn@doi [MNRAS]
  {10.1046/j.1365-8711.2002.05038.x}, 329, 897

\bibitem[\protect\citeauthoryear{Iben \& Tutukov}{Iben \&
  Tutukov}{1996}]{iben_origin_1996}
Iben I.,  Tutukov A.~V.,  1996, \mn@doi [ApJ] {10.1086/176693}, 456, 738

\bibitem[\protect\citeauthoryear{Ivanova, Belczynski, Kalogera, Rasio  \&
  Taam}{Ivanova et~al.}{2003}]{ivanova_role_2003}
Ivanova N.,  Belczynski K.,  Kalogera V.,  Rasio F.~A.,   Taam R.~E.,  2003,
  \mn@doi [ApJ] {10.1086/375578}, 592, 475

\bibitem[\protect\citeauthoryear{Janka}{Janka}{2013}]{janka_natal_2013}
Janka H.-T.,  2013, \mn@doi [MNRAS] {10.1093/mnras/stt1106}, 434, 1355

\bibitem[\protect\citeauthoryear{Janka}{Janka}{2017}]{janka_neutron_2017}
Janka H.-T.,  2017, \mn@doi [ApJ] {10.3847/1538-4357/aa618e}, 837, 84

\bibitem[\protect\citeauthoryear{Janka \& M{\"u}ller}{Janka \&
  M{\"u}ller}{1994}]{janka_neutron_1994}
Janka H.-T.,  M{\"u}ller E.,  1994, A\&A, 290, 496

\bibitem[\protect\citeauthoryear{Kalogera}{Kalogera}{1996}]{kalogera_orbital_1996}
Kalogera V.,  1996, \mn@doi [ApJ] {10.1086/177974}, 471, 352

\bibitem[\protect\citeauthoryear{Kasliwal et~al.,}{Kasliwal
  et~al.}{2017}]{kasliwal_illuminating_2017}
Kasliwal M.~M.,  et~al., 2017, \mn@doi [Science] {10.1126/science.aap9455},
  358, 1559

\bibitem[\protect\citeauthoryear{Katsuda et~al.,}{Katsuda
  et~al.}{2018}]{katsuda_intermediate-mass_2018}
Katsuda S.,  et~al., 2018, \mn@doi [ApJ] {10.3847/1538-4357/aab092}, 856, 18

\bibitem[\protect\citeauthoryear{Kochanek, Auchettl  \& Belczynski}{Kochanek
  et~al.}{2019}]{kochanek_stellar_2019}
Kochanek C.~S.,  Auchettl K.,   Belczynski K.,  2019, \mn@doi [MNRAS]
  {10.1093/mnras/stz717}, 485, 5394

\bibitem[\protect\citeauthoryear{Kruckow, Tauris, Langer, Kramer  \&
  Izzard}{Kruckow et~al.}{2018}]{kruckow_progenitors_2018}
Kruckow M.~U.,  Tauris T.~M.,  Langer N.,  Kramer M.,   Izzard R.~G.,  2018,
  \mn@doi [MNRAS] {10.1093/mnras/sty2190}, 481, 1908

\bibitem[\protect\citeauthoryear{Kusenko}{Kusenko}{2004}]{kusenko_pulsar_2004}
Kusenko A.,  2004, \mn@doi [IJMPD] {10.1142/S0218271804006486}, 13, 2065

\bibitem[\protect\citeauthoryear{Kusenko \& Segr\`{e}}{Kusenko \&
  Segr\`{e}}{1996}]{kusenko_pulsar_1996}
Kusenko A.,  Segr\`{e} G.,  1996, \mn@doi [PRL] {10.1103/PhysRevLett.77.4872},
  77, 4872

\bibitem[\protect\citeauthoryear{{LIGO-Virgo Collaboration}}{{LIGO-Virgo
  Collaboration}}{2016a}]{ligo_scientific_collaboration_and_virgo_collaboration_observation_2016}
{LIGO-Virgo Collaboration} 2016a, \mn@doi [PRL]
  {10.1103/PhysRevLett.116.061102}, 116, 061102

\bibitem[\protect\citeauthoryear{{LIGO-Virgo Collaboration}}{{LIGO-Virgo
  Collaboration}}{2016b}]{ligo_scientific_collaboration_and_virgo_collaboration_gw151226:_2016}
{LIGO-Virgo Collaboration} 2016b, \mn@doi [PRL]
  {10.1103/PhysRevLett.116.241103}, 116, 241103

\bibitem[\protect\citeauthoryear{{LIGO-Virgo Collaboration}}{{LIGO-Virgo
  Collaboration}}{2017a}]{abbott_exploring_2017}
{LIGO-Virgo Collaboration} 2017a, \mn@doi [CQG] {10.1088/1361-6382/aa51f4}, 34,
  044001

\bibitem[\protect\citeauthoryear{{LIGO-Virgo Collaboration}}{{LIGO-Virgo
  Collaboration}}{2017b}]{abbott_gw170817:_2017}
{LIGO-Virgo Collaboration} 2017b, \mn@doi [PRL]
  {10.1103/PhysRevLett.119.161101}, 119, 161101

\bibitem[\protect\citeauthoryear{{LIGO-Virgo Collaboration}}{{LIGO-Virgo
  Collaboration}}{2017c}]{abbott_progenitor_2017}
{LIGO-Virgo Collaboration} 2017c, \mn@doi [ApJL] {10.3847/2041-8213/aa93fc},
  850, L40

\bibitem[\protect\citeauthoryear{MacLeod \& Ramirez-Ruiz}{MacLeod \&
  Ramirez-Ruiz}{2014}]{macleod_accretion-fed_2014}
MacLeod M.,  Ramirez-Ruiz E.,  2014, \mn@doi [ApJL]
  {10.1088/2041-8205/798/1/L19}, 798, L19

\bibitem[\protect\citeauthoryear{Macleod, Harry  \& Fairhurst}{Macleod
  et~al.}{2016}]{macleod_fully-coherent_2016}
Macleod D.,  Harry I.,   Fairhurst S.,  2016, \mn@doi [PRD]
  {10.1103/PhysRevD.93.064004}, 93, 064004

\bibitem[\protect\citeauthoryear{Mandel}{Mandel}{2016}]{mandel_estimates_2016}
Mandel I.,  2016, \mn@doi [MNRAS] {10.1093/mnras/stv2733}, 456, 578

\bibitem[\protect\citeauthoryear{Mapelli \& Giacobbo}{Mapelli \&
  Giacobbo}{2018}]{mapelli_cosmic_2018}
Mapelli M.,  Giacobbo N.,  2018, \mn@doi [MNRAS] {10.1093/mnras/sty1613}, 479,
  4391

\bibitem[\protect\citeauthoryear{Martin, Tout  \& Pringle}{Martin
  et~al.}{2009}]{martin_supernova_2009}
Martin R.~G.,  Tout C.~A.,   Pringle J.~E.,  2009, \mn@doi [MNRAS]
  {10.1111/j.1365-2966.2009.15031.x}, 397, 1563

\bibitem[\protect\citeauthoryear{Michaely \& Perets}{Michaely \&
  Perets}{2018}]{michaely_supernova_2018}
Michaely E.,  Perets H.~B.,  2018, \mn@doi [ApJL] {10.3847/2041-8213/aaacfc},
  855, L12

\bibitem[\protect\citeauthoryear{Mink \& Belczynski}{Mink \&
  Belczynski}{2015}]{mink_merger_2015}
Mink S. E.~d.,  Belczynski K.,  2015, \mn@doi [ApJ]
  {10.1088/0004-637X/814/1/58}, 814, 58

\bibitem[\protect\citeauthoryear{Morozova, Radice, Burrows  \&
  Vartanyan}{Morozova et~al.}{2018}]{morozova_gravitational_2018}
Morozova V.,  Radice D.,  Burrows A.,   Vartanyan D.,  2018, \mn@doi [ApJ]
  {10.3847/1538-4357/aac5f1}, 861, 10

\bibitem[\protect\citeauthoryear{M\"{u}ller, Gay, Heger, Tauris  \&
  Sim}{M\"{u}ller et~al.}{2018}]{muller_multidimensional_2018}
M\"{u}ller B.,  Gay D.~W.,  Heger A.,  Tauris T.~M.,   Sim S.~A.,  2018,
  \mn@doi [MNRAS] {10.1093/mnras/sty1683}, 479, 3675

\bibitem[\protect\citeauthoryear{M\"{u}ller et~al.,}{M\"{u}ller
  et~al.}{2019}]{muller_three-dimensional_2019}
M\"{u}ller B.,  et~al., 2019, \mn@doi [MNRAS] {10.1093/mnras/stz216}, 484, 3307

\bibitem[\protect\citeauthoryear{Nakamura, Takiwaki  \& Kotake}{Nakamura
  et~al.}{2019}]{nakamura_long-term_2019}
Nakamura K.,  Takiwaki T.,   Kotake K.,  2019, PASJ, submitted
  (arXiv:1904.08088)

\bibitem[\protect\citeauthoryear{Nordhaus, Brandt, Burrows, Livne  \&
  Ott}{Nordhaus et~al.}{2010}]{nordhaus_theoretical_2010}
Nordhaus J.,  Brandt T.~D.,  Burrows A.,  Livne E.,   Ott C.~D.,  2010, \mn@doi
  [PRD] {10.1103/PhysRevD.82.103016}, 82, 103016

\bibitem[\protect\citeauthoryear{Nordhaus, Brandt, Burrows  \&
  Almgren}{Nordhaus et~al.}{2012}]{nordhaus_hydrodynamic_2012}
Nordhaus J.,  Brandt T.~D.,  Burrows A.,   Almgren A.,  2012, \mn@doi [MNRAS]
  {10.1111/j.1365-2966.2012.21002.x}, 423, 1805

\bibitem[\protect\citeauthoryear{O'Shaughnessy, Kim, Fragos, Kalogera  \&
  Belczy\'{n}ski}{O'Shaughnessy et~al.}{2005}]{oshaughnessy_constraining_2005}
O'Shaughnessy R.,  Kim C.,  Fragos T.,  Kalogera V.,   Belczy\'{n}ski K.,
  2005, \mn@doi [ApJ] {10.1086/468180}, 633, 1076

\bibitem[\protect\citeauthoryear{O'Shaughnessy, Gerosa  \&
  Wysocki}{O'Shaughnessy et~al.}{2017}]{oshaughnessy_inferences_2017}
O'Shaughnessy R.,  Gerosa D.,   Wysocki D.,  2017, \mn@doi [PRL]
  {10.1103/PhysRevLett.119.011101}, 119

\bibitem[\protect\citeauthoryear{Os\l{}owski, Bulik, Gondek-Rosi\'{n}ska  \&
  Belczy\'{n}ski}{Os\l{}owski et~al.}{2011}]{oslowski_population_2011}
Os\l{}owski S.,  Bulik T.,  Gondek-Rosi\'{n}ska D.,   Belczy\'{n}ski K.,  2011,
  \mn@doi [MNRAS] {10.1111/j.1365-2966.2010.18147.x}, 413, 461

\bibitem[\protect\citeauthoryear{{Pajkos}, {Couch}, {Pan}  \&
  {O{\textquoteright}Connor}}{{Pajkos} et~al.}{2019}]{pajkos_features_2019}
{Pajkos} M.~A.,  {Couch} S.~M.,  {Pan} K.-C.,   {O{\textquoteright}Connor}
  E.~P.,  2019, \mn@doi [\apj] {10.3847/1538-4357/ab1de2}, \href
  {https://ui.adsabs.harvard.edu/abs/2019ApJ...878...13P} {878, 13}

\bibitem[\protect\citeauthoryear{Postnov \& Yungelson}{Postnov \&
  Yungelson}{2006}]{postnov_evolution_2006}
Postnov K.~A.,  Yungelson L.~R.,  2006, \mn@doi [Living Rev. Relativ.]
  {10.12942/lrr-2006-6}, 9, 6

\bibitem[\protect\citeauthoryear{Powell \& M\"{u}ller}{Powell \&
  M\"{u}ller}{2019}]{powell_gravitational_2019}
Powell J.,  M\"{u}ller B.,  2019, \mn@doi [MNRAS] {10.1093/mnras/stz1304}, 487,
  1178

\bibitem[\protect\citeauthoryear{Powell, Gossan, Logue  \& Heng}{Powell
  et~al.}{2016}]{powell_inferring_2016}
Powell J.,  Gossan S.~E.,  Logue J.,   Heng I.~S.,  2016, \mn@doi [PRD]
  {10.1103/PhysRevD.94.123012}, 94, 123012

\bibitem[\protect\citeauthoryear{Pradier, Arnaud, Bizouard, Cavalier, Davier
  \& Hello}{Pradier et~al.}{2001}]{pradier_efficient_2001}
Pradier T.,  Arnaud N.,  Bizouard M.-A.,  Cavalier F.,  Davier M.,   Hello P.,
  2001, \mn@doi [PRD] {10.1103/PhysRevD.63.042002}, 63, 042002

\bibitem[\protect\citeauthoryear{Radice, Morozova, Burrows, Vartanyan  \&
  Nagakura}{Radice et~al.}{2019}]{radice_characterizing_2019}
Radice D.,  Morozova V.,  Burrows A.,  Vartanyan D.,   Nagakura H.,  2019,
  \mn@doi [ApJL] {10.3847/2041-8213/ab191a}, 876, L9

\bibitem[\protect\citeauthoryear{Repetto, Davies  \& Sigurdsson}{Repetto
  et~al.}{2012}]{repetto_investigating_2012}
Repetto S.,  Davies M.~B.,   Sigurdsson S.,  2012, \mn@doi [MNRAS]
  {10.1111/j.1365-2966.2012.21549.x}, 425, 2799

\bibitem[\protect\citeauthoryear{Roma, Powell, Heng  \& Frey}{Roma
  et~al.}{2019}]{roma_astrophysics_2019}
Roma V.,  Powell J.,  Heng I.~S.,   Frey R.,  2019, \mn@doi [PRD]
  {10.1103/PhysRevD.99.063018}, 99, 063018

\bibitem[\protect\citeauthoryear{Sagert \& Schaffner-Bielich}{Sagert \&
  Schaffner-Bielich}{2008}]{sagert_pulsar_2008}
Sagert I.,  Schaffner-Bielich J.,  2008, \mn@doi [A\&A]
  {10.1051/0004-6361:20078530}, 489, 281

\bibitem[\protect\citeauthoryear{Sato et~al.,}{Sato
  et~al.}{2009}]{sato_decigo:_2009}
Sato S.,  et~al., 2009, \mn@doi [J. Phys.: Conf. Ser.]
  {10.1088/1742-6596/154/1/012040}, 154, 012040

\bibitem[\protect\citeauthoryear{Sato et~al.,}{Sato
  et~al.}{2017}]{sato_status_2017}
Sato S.,  et~al., 2017, \mn@doi [J. Phys.: Conf. Ser.]
  {10.1088/1742-6596/840/1/012010}, 840, 012010

\bibitem[\protect\citeauthoryear{Scheck, Plewa, Janka, Kifonidis  \&
  M\"{u}ller}{Scheck et~al.}{2004}]{scheck_pulsar_2004}
Scheck L.,  Plewa T.,  Janka H.-T.,  Kifonidis K.,   M\"{u}ller E.,  2004,
  \mn@doi [PRL] {10.1103/PhysRevLett.92.011103}, 92, 011103

\bibitem[\protect\citeauthoryear{Scheck, Kifonidis, Janka  \&
  M\"{u}ller}{Scheck et~al.}{2006}]{scheck_multidimensional_2006}
Scheck L.,  Kifonidis K.,  Janka H.-T.,   M\"{u}ller E.,  2006, \mn@doi [A\&A]
  {10.1051/0004-6361:20064855}, 457, 963

\bibitem[\protect\citeauthoryear{Schinzel, Kerr, Rau, Bhatnagar  \&
  Frail}{Schinzel et~al.}{2019}]{schinzel_tail_2019}
Schinzel F.~K.,  Kerr M.,  Rau U.,  Bhatnagar S.,   Frail D.~A.,  2019, \mn@doi
  [ApJL] {10.3847/2041-8213/ab18f7}, 876, L17

\bibitem[\protect\citeauthoryear{Socrates, Blaes, Hungerford  \&
  Fryer}{Socrates et~al.}{2005}]{socrates_neutrino_2005}
Socrates A.,  Blaes O.,  Hungerford A.,   Fryer C.~L.,  2005, \mn@doi [ApJ]
  {10.1086/431786}, 632, 531

\bibitem[\protect\citeauthoryear{Tauris, Langer, Moriya, Podsiadlowski, Yoon
  \& Blinnikov}{Tauris et~al.}{2013}]{tauris_ultra-stripped_2013}
Tauris T.~M.,  Langer N.,  Moriya T.~J.,  Podsiadlowski P.,  Yoon S.-C.,
  Blinnikov S.~I.,  2013, \mn@doi [ApJL] {10.1088/2041-8205/778/2/L23}, 778,
  L23

\bibitem[\protect\citeauthoryear{Tauris, Langer  \& Podsiadlowski}{Tauris
  et~al.}{2015}]{tauris_ultra-stripped_2015}
Tauris T.~M.,  Langer N.,   Podsiadlowski P.,  2015, \mn@doi [MNRAS]
  {10.1093/mnras/stv990}, 451, 2123

\bibitem[\protect\citeauthoryear{Tauris et~al.,}{Tauris
  et~al.}{2017}]{tauris_formation_2017}
Tauris T.~M.,  et~al., 2017, \mn@doi [ApJ] {10.3847/1538-4357/aa7e89}, 846, 170

\bibitem[\protect\citeauthoryear{Taylor \& Gerosa}{Taylor \&
  Gerosa}{2018}]{taylor_mining_2018}
Taylor S.~R.,  Gerosa D.,  2018, \mn@doi [PRD] {10.1103/PhysRevD.98.083017},
  98, 083017

\bibitem[\protect\citeauthoryear{Tzanavaris et~al.,}{Tzanavaris
  et~al.}{2013}]{tzanavaris_modeling_2013}
Tzanavaris P.,  et~al., 2013, \mn@doi [ApJ] {10.1088/0004-637X/774/2/136}, 774,
  136

\bibitem[\protect\citeauthoryear{Vigna-G\'{o}mez et~al.,}{Vigna-G\'{o}mez
  et~al.}{2018}]{vigna-gomez_formation_2018}
Vigna-G\'{o}mez A.,  et~al., 2018, \mn@doi [MNRAS] {10.1093/mnras/sty2463},
  481, 4009

\bibitem[\protect\citeauthoryear{Wang, Lai  \& Han}{Wang
  et~al.}{2006}]{wang_neutron_2006}
Wang C.,  Lai D.,   Han J.~L.,  2006, \mn@doi [ApJ] {10.1086/499397}, 639, 1007

\bibitem[\protect\citeauthoryear{Wheeler, Lecar  \& McKee}{Wheeler
  et~al.}{1975}]{wheeler_supernovae_1975}
Wheeler J.~C.,  Lecar M.,   McKee C.~F.,  1975, \mn@doi [ApJ] {10.1086/153771},
  200, 145

\bibitem[\protect\citeauthoryear{Willems, Kalogera  \& Henninger}{Willems
  et~al.}{2004}]{willems_pulsar_2004}
Willems B.,  Kalogera V.,   Henninger M.,  2004, \mn@doi [ApJ]
  {10.1086/424812}, 616, 414

\bibitem[\protect\citeauthoryear{Wong, Willems  \& Kalogera}{Wong
  et~al.}{2010}]{wong_constraints_2010}
Wong T.-W.,  Willems B.,   Kalogera V.,  2010, \mn@doi [ApJ]
  {10.1088/0004-637X/721/2/1689}, 721, 1689

\bibitem[\protect\citeauthoryear{Wongwathanarat, Janka  \&
  M\"{u}ller}{Wongwathanarat et~al.}{2010}]{wongwathanarat_hydrodynamical_2010}
Wongwathanarat A.,  Janka H.-T.,   M\"{u}ller E.,  2010, \mn@doi [ApJL]
  {10.1088/2041-8205/725/1/L106}, 725, L106

\bibitem[\protect\citeauthoryear{Wongwathanarat, Janka  \&
  M\"{u}ller}{Wongwathanarat
  et~al.}{2013}]{wongwathanarat_three-dimensional_2013}
Wongwathanarat A.,  Janka H.-T.,   M\"{u}ller E.,  2013, \mn@doi [A\&A]
  {10.1051/0004-6361/201220636}, 552, A126

\bibitem[\protect\citeauthoryear{Wysocki, Gerosa, O'Shaughnessy, Belczynski,
  Gladysz, Berti, Kesden  \& Holz}{Wysocki
  et~al.}{2018}]{wysocki_explaining_2018}
Wysocki D.,  Gerosa D.,  O'Shaughnessy R.,  Belczynski K.,  Gladysz W.,  Berti
  E.,  Kesden M.,   Holz D.~E.,  2018, \mn@doi [PRD]
  {10.1103/PhysRevD.97.043014}, 97, 043014

\bibitem[\protect\citeauthoryear{Yagi \& Seto}{Yagi \&
  Seto}{2011}]{yagi_detector_2011}
Yagi K.,  Seto N.,  2011, \mn@doi [PRD] {10.1103/PhysRevD.83.044011}, 83,
  044011

\makeatother
\end{thebibliography}

%%%%%%%%%%%%%%%%%%%%%%%%%%%%%%%%%%%%%%%%%%%%%%%%%%

%%%%%%%%%%%%%%%%% APPENDICES %%%%%%%%%%%%%%%%%%%%%

\appendix
%\section{Journal abbreviations}

%%%%%%%%%%%%%%%%%%%%%%%%%%%%%%%%%%%%%%%%%%%%%%%%%%

% Don't change these lines
\bsp	% typesetting comment
\label{lastpage}
\end{document}